\documentclass[10pt,preprint,a4paper]{aastex}

\usepackage{amsmath}                % American Mathematical Society package
\usepackage{amsfonts}               % American Mathematical Society fonts
\usepackage{amssymb}                % American Mathematical Society symbol
\usepackage{epsfig}                 % EPS figures

\newcommand{\kms}{{~\rm km\; s^{-1}}}

\newcommand{\cm}{{~\rm cm}}
\newcommand{\s}{{~\rm s}}

\newcommand{\K}{{~\rm K}}
\newcommand{\erg}{{~\rm erg}}
\newcommand{\yr}{{~\rm yr}}

\newcommand{\pc}{{~\rm pc}}
\newcommand{\kpc}{{~\rm kpc}}

\begin{document}

\title{SUPPRESSING HOT GAS ACCRETION TO SUPERMASSIVE BLACK HOLES BY STELLAR WINDS}

\author{Shlomi Hillel\altaffilmark{1} and Noam Soker\altaffilmark{1}}

\altaffiltext{1}{Department of Physics, Technion -- Israel Institute of Technology, Haifa
32000, Israel; shlomihi@tx.technion.ac.il, soker@physics.technion.ac.il}

\begin{abstract}
We argue that one of the basic assumptions of the Bondi accretion process, that the accreting object has zero pressure,
might not hold in many galaxies because of the pressure exerted by stellar winds of star orbiting the central super massive black hole (SMBH).
Hence, the Bondi accretion cannot be used in these cases, such as in the galaxy NGC~3115.
The winds of these high-velocity stars are shocked to temperatures above the virial temperature of the galaxy, leading to the formation of
a hot bubble of size $\sim 0.1-10 \pc$ near the center. This hot bubble can substantially reduce the mass accretion rate by the SMBH.
If the density of the hot bubble is lower than that of the interstellar medium (ISM), a density-inversion layer is formed.
{{{ As the gas loses energy by X-ray radiation, eventually more mass of the cooling shocked stellar winds will be accreted to the
SMBH. This accretion will be of cold clumps.
After a period of millions of years of low AGN activity, therefore, a stronger AGN activity will occur that will heat and expel gas,
much as in cooling flow clusters. }}}
Adding to other problems of the Bondi process, our results render the Bondi accretion irrelevant
for AGN feedback in cooling flow in galaxies and small groups of galaxies and during galaxy formation.
\end{abstract}

%%% \keywords{ galaxies: clusters: intracluster medium, galaxies: clusters: cooling flows, galaxies: clusters: general  galaxies: individual: NGC3115 }

% ==========================================================
\section{INTRODUCTION}
\label{s-intro}
% ==========================================================

It is widely accepted that feedback powered by active galactic nuclei (AGN) has a key role in galaxy formation and
in cooling flows in galaxies and in clusters of galaxies.
In galaxy formation AGN feedback heats and expels gas {{{ from the galaxy}}} (e.g., \citealt{Bower2008, Ostriker2010} and references therein),
and by that can determine the correlation between the central super-massive black hole (SMBH) mass and some
properties of the galaxy {{{ \citep{King2003,King2005,Soker2009b,Soker2011}}}}.
In cooling flow clusters jets launched by the SMBH heat the gas and maintain
a small, but non zero cooling flow (see review by \citealt{McNamara2007, McNamara2012, Fabian2012});
this is termed a moderate cooling flow.

There is a dispute on how the accretion onto the SMBH occurs, in particular in cooling flows.
One camp argues for accretion to be of hot gas via the Bondi accretion process
(e.g., \citealp{Allen2006, Russell2010, Narayan2011}),
while the other side argues that the accretion is of dense and
cold clumps in what is termed the cold feedback mechanism {{{ \citep{Pizzolato2005,Pizzolato2010}}}}.
The cold feedback mechanism has been strengthened recently by observations
of cold gas and by more detailed studies \citep{Revaz2008,Pope2009,Wilman2009,Wilman2011,Nesvadba2011,Cavagnolo2011,Gaspari2012a,Gaspari2012b,
McCourt2012,Sharma2012,Farage2012,Kashi2012}.

The Bondi accretion process, on the other hand, suffers from two problems.
The first problem is that in cooling-flow clusters the Bondi accretion rate
is too low to account for the AGN power
(e.g., \citealt{McNamara2011, Cavagnolo2011}).
The second is that there is no time for the feedback to work \citep{Soker2009}.
{{{ This is because the time for cooling gas at distances of $\ga {\rm few} \times \kpc$
in the Bondi accretion process to be accreted and power jets
that heat back the ISM, is much longer than the cooling time of the gas. }}}
This is already true for gas cooling at a moderate distance of $\sim 1 \kpc$ from the center.
In other words the gas at large distances has no time to communicate with the SMBH before it cools.

In this paper we point out yet another problematic point with the Bondi accretion process.
In a recent paper, \citet{Wong2011} resolved the region within the Bondi accretion radius of the S0 galaxy NGC~3115.
If the density and temperature profile is interpreted as resulting from a Bondi
accretion flow onto the $M_{\rm BH}=2 \times 10^9 M_\odot$ central SMBH, the derived accretion rate is
$\dot M_B=2.2 \times 10^{-2} M_\odot \yr^{-1}$.
They note that for a radiation power of $0.1 \dot{M_B}\, c^2$, the expected accretion luminosity is
six orders of magnitude above the observed upper limit.
They attribute this to a process where most of the inflowing gas is blown away,
{{{ or the gas is continuously circulating in convective eddies,
or to that the region they resolve is not yet incorporated to the Bondi accretion flow.
The idea of circulating eddies has some similarities to the density inversion layer behavior we discuss here. }}}

{{{ In any case, some AGN activity does take place in NGC~3115 \citep{Wrobel2012}.
\cite{Wrobel2012} detected a radio nucleus in NGC~3115 with a radio power of $L_{\rm radio}=3 \times 10^{35} \erg \s^{-1}$.
This indicates the presence of a weak AGN, that might substantially reduce the accretion rate \citep{Wrobel2012}.
As we discuss later, the feeding of the SMBH might be from the stellar winds rather than from the ISM. }}}

{{{ Several other processes were considered to reduce the accretion rate by a SMBH much below the Bondi accretion rate.
Such processes include magnetic field reconnection \citep{Igumenshchev2002}, angular momentum \citep{Proga2003a,Proga2003b},
magneto-thermal instabilities \citep{Sharma2008}, and instabilities due to self-gravitation of the infalling gas \citep{Levine2010}.
Lack of spherical symmetry in realistic situations is an additional factor \citep{Debuhr2011}.
Turbulent media can have higher than Bondi-Hoyle accretion rate, but due to vorticity, a lower
accretion rate is also possible \citep{Krumholz2005,Krumholz2006}.
\cite{Hobbs2012} claim that the Bondi-Hoyle solution is only relevant for hot
virialized gas with no angular momentum and negligible radiative cooling.
}}}

We take a different view
{{{ on the suppression of the Bondi accretion.
We argue that in many galaxies for a fraction of the time the Bondi accretion flow
might not be relevant because }}}
one cannot assume a zero pressure at the center, either because of
stellar winds or because of jets blown by the AGN.

% ==========================================================
\section{THE PRESSURE OF STELLAR WINDS}
\label{s-ramstars}
% ==========================================================

The pressure exerted by stellar winds of high velocity stars (i.e., moving much faster than the dispersion velocity in the galaxy)
with an average mass loss rate per star of $\dot m_\ast$ can be calculated in two limits, which basically lead to the same result.
First we calculate the pressure by considering the total outward momentum flux at radius $r$.
{{{ Because the orbital velocities of stars around the SMBH are much larger than the typical
velocities of the stellar winds (as most of the mass loss is during the asymptotic giant branch, AGB, phase), }}}
the relevant velocity in general is not that of the wind relative to the star, but
rather the velocity of the star under the gravitational influence of the SMBH,
\begin{equation}
u_\ast (r) \simeq \sqrt{\frac{G M_{\rm BH}}{r}} = 2 \times 10^3   % 2.07
\left( \frac{M_{\rm BH}}{10^9 M_\odot} \right) ^ {1/2}
\left( \frac{r}{\pc} \right) ^ {-1/2} \kms.
\label{eq:vbh1}
\end{equation}
This holds as long as the SMBH gravity dominates that of the galaxy.
In NGC~3115 that we study in more detail in section \ref{s-ngc3115}, for example, the SMBH gravity dominates that
of the galaxy to a distance of $\sim 30 \pc$ as the black hole mass is $M_{\rm BH}=2 \times 10^9 M_\odot$.
Let stellar winds from high-velocity stars dominate the pressure inside a sphere of radius $R_h$.
The pressure exerted by the wind on a surface of radius $R_h$ is approximately given by adding the ram pressures of winds from  all stars inside the sphere of radius $R_h$,
\begin{equation}
P_{m\ast} (R_h) \simeq n_\ast \eta \dot m_\ast u_\ast (R_h) \frac{4 \pi R_h^3}{3} \frac{1}{4 \pi R_h^2},
\label{eq:ramp1}
\end{equation}
where $n_\ast$ is the stellar {{{ number}}} density in the center of the galaxy, and $\eta$ is the fraction of the mass
lost by stars that is shocked and heats up.
In all our expressions the stellar mass loss rate appears as $\eta \dot m_\ast$.

Some of the mass lost by stars will form dense clumps that will cool rapidly even
if being shocked, or will not even be shocked.
This is particularly true as most of the mass is being lost by AGB stars that have dense winds.
The thermal pressure of the ISM in the center will cause part of the winds' gas to form dense clouds.
{{{ Many of the cold clumps can be evaporated by heat conduction form the hot gas in the bubble.
However, some clumps might flow inward and feed the SMBH, and explain the AGN activity observed by \cite{Wrobel2012}. }}}
The average mass loss rate is calculated as follows. A solar-like star loses $\sim 0.5 M_\odot$ over $\sim 10^{10} \yr$.
Considering an old population of stars, the mass loss rate is lower even.
More accurately, most of the mass loss is due to AGB stars, which live for $\sim 10^7 \yr$, and lose mass
at an average rate of $\sim 10^{-7} M_\odot \yr^{-1}$ \citep{Willson2007}.
During the final stages of the AGB the evolution is faster and the mass loss rate is higher.
If there is a young stellar population, the total mass loss rate can be much higher.
The ram pressure will not increase much beyond few~pc because the stellar density decreases.

%%%   \begin{equation}
%%%   \begin{split}
%%%   P_{m\ast} (r) &
%%%   \simeq n_\ast \dot m_\ast (r) u_\ast \frac{4 \pi r^3}{3} \frac{1}{4 \pi r^2}
%%%   \\& = 2.3 \times 10^{-8}
%%%   \left( \frac {M_{\rm BH}} {10^9 M_\odot} \right)^{1/2}
%%%   \left( \frac {r} {1 \pc} \right)^{1/2}
%%%   \left( \frac {n_\ast (r)} {5\times 10^5 \pc^{-3}} \right)
%%%   \left( \frac {\dot m_\ast } {10^{-10} M_\odot \yr^{-1} } \right) \erg \cm^{-3},
%%%   \end{split}
%%%   \label{eq:pe1}
%%%   \end{equation}

An alternative point of view would be to express the pressure as (roughly) the energy density of the shocked stellar wind.
We also assume a constant pressure and density inside this sphere.
{{{ This is justified because we are interested mainly in the outer part of the hot bubble,
where density inversion might take place. Even a steep power law profile, say of $\rho \sim r^{-2}$,
will not change much the density from  $0.5 R_h$ to $R_h$, which contains 0.875 of the volume of the bubble. }}}
We can calculate the rate of energy input and multiply by the time it takes the hot gas to leave the inner region
\begin{equation}
\tau_{\rm esc} = \frac{R_h}{\beta u_\ast (R_h)},
\label{eq:pe11}
\end{equation}
where $\beta \la 1$ takes into consideration that the hot gas at the center escapes at velocity
lower than the escape velocity.
The stellar wind pressure in this case can be written as
\begin{equation}
P_{e\ast} = \frac{2}{3} \frac{\dot{E}}{V}\frac{R_h}{\beta u_\ast (R_h)},
\label{eq:pe1}
\end{equation}
where the energy deposition rate is
\begin{equation}
\dot{E}=\int_0^{R_h}{\left(\frac{1}{2} n_\ast \eta \dot{m_\ast} u_\ast^2 (r) \right) 4\pi r^2 dr} =
2 \pi G M_{\rm BH} \eta \dot{m_\ast} \int_0^{R_h} {n_\ast(r) r dr}.
\label{eq:e1}
\end{equation}
Scaling the different quantities and assuming a constant stellar density we find
\begin{equation}
P_{e\ast}
= 3 \times 10^{-8} \beta^{-1} \eta
\left( \frac {M_{\rm BH}} {10^9 M_\odot} \right)^{1/2}
\left( \frac {R_h} {1 \pc} \right)^{1/2}
\left( \frac {n_\ast} {5\times 10^5 \pc^{-3}} \right)
\left( \frac {\dot m_\ast } {10^{-10} M_\odot \yr^{-1} } \right) \erg \cm^{-3},
\label{eq:pe2}
\end{equation}
where the stellar density is scaled by the average stellar density within $\sim 3 \pc$ from
the center of NGC~3115 \citep{Kormendy1996}.
Equations (\ref{eq:pe1}) and  (\ref{eq:pe2}) are more accurate than equation (\ref{eq:ramp1}) when the radiative cooling time
of the colliding stellar winds is larger than the escape time $\tau_{\rm esc}$, which is the case here due to the high-temperature low-density
post-shock stellar winds.
{{{ The radiative cooling time is $\tau_c = (5/2) nk T/(n_e n_ p \Lambda) \simeq 10^7-10^8$ years,
This is much longer than the escape time given in equation (\ref{eq:pe11}) $\tau_{\rm esc} \simeq 10^2-10^3$ years.
Here $n_e$, $n_p$, and $n$ are the electron, proton, and total number density, respectively, and $\Lambda$ is the
cooling function.  }}}
Therefore, from now on we will refer to the hot gas region formed by the shocked stellar winds as the hot bubble,
{{{ and to its radius as $R_h$. }}}
For a constant stellar density within radius $r$, we find $P_{e\ast} = \frac{3}{2} \beta^{-1} P_{m\ast}$.
If the stellar density drops to zero at some radius $r_z$ (a nonrealistic ideal case), the pressure beyond $r_z$
will drop like $(r/r_z)^{-2}$.

\begin{figure}[htb]
\begin{center}
\includegraphics[width=0.75\textwidth]{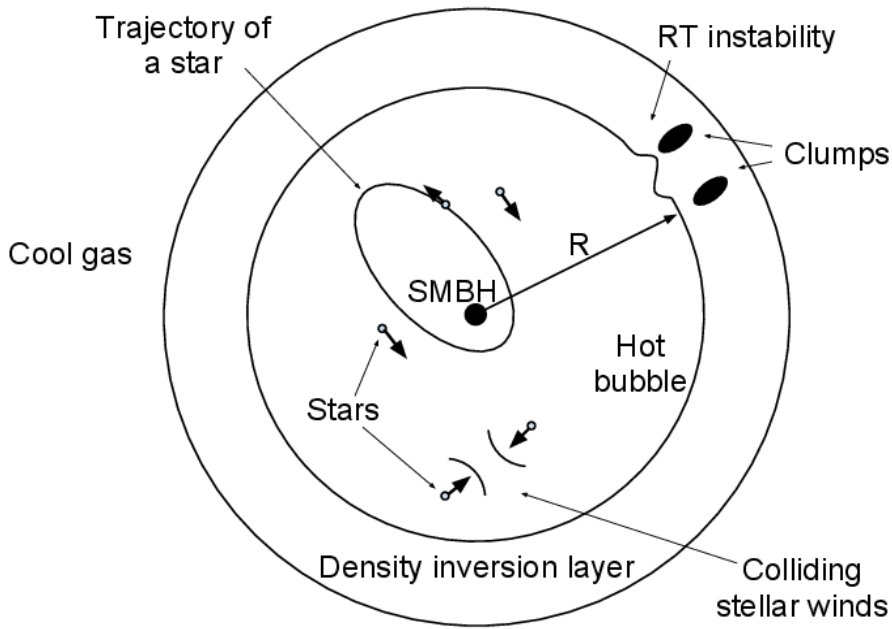}
%%% \hspace{1.5in}\parbox{6in}{\caption{A schematic drawing (not to scale) of the flow structure.} }
\caption{A schematic drawing (not to scale) of the flow structure where a hot bubble, formed by stellar winds of high-velocity stars orbiting the
central SMBH, exerts pressure on the ISM residing outside radius $R_h$.
If the density in the hot bubble is lower than the ISM density, the flow at $R_h$ is RT-unstable and a density-inversion layer is formed.
{{{ Most clumps that are formed in the winds collision process are later evaporated by heat conduction from the hot bubble to
the clumps. Some, thought.
are accreted by the SMBH and explain the weak AGN activity observed by \cite{Wrobel2012}. }}}    }
\label{fig:fig1}
\end{center}
\end{figure}

The average density of the hot shocked stellar wind is given by
\begin{equation}
\rho_w \simeq  \left( \frac{4 \pi}{3} R^3_h \right)^{-1} \eta \dot{m_\ast} \frac{R_h}{\beta u_\ast(R_h)}
\int_0^{R_h} 4 \pi {n_\ast(r) r^2 dr}
\label{eq:rho1}
\end{equation}
The flow structure is schematically drawn in Fig. \ref{fig:fig1}.
{{{ Relevant to this flow structure is the simulations of \cite{Cuadra2008}. They simulated the dynamics of stellar
winds in the Galactic center and found the accretion rate to be highly variable, due in part to the stochastic nature of infalling cold clumps.
\cite{Fryer2007} suggest that the inner $\sim 5 \pc$ region surrounding Sgr A$^{\ast}$ in our Galaxy
can be approximated by a wind-blown hot bubble density structure.
}}}
% FFFFFFFFFFFFFFFFFFFFFFFFFFFFFFFFFFFFFFFFFFFFFFFF
% FFFFFFFFFFFFFFFFFFFFFFFFFFFFFFFFFFFFFFFFFFFFFFFF
%
%    Kormendy et al. 1996 give r_h=2pc with escape speed of 352 km/sec. This gives a mass of 2.9x10^7Mo.
%    The density is 8.6e5 Mo pc^{-3}. Else, they give L_v=1.2e7 and M/L = 3.1 which gives M=3.72e7 inside
%    the half light radius of 2pc. This gives a mass of ~1.5e7 inside 2pc, or a density of ~10^6 Mo pc^{-3}.
%    They also give cross time of 15,800 years, which comes from (G \rho)^{-1/2}=1.57year
%    for \rho = 9e5 Mo pc^{-3}.

% ==========================================================
\section{THE CASE OF NGC~3115}
\label{s-ngc3115}
% ==========================================================

At the Bondi radius $R_B \simeq 210 \pc$ of the galaxy NGC~3115 the ISM pressure is
$P(R_B)=2 \times 10^{-11} \erg \cm^{-3}$, the electron number density is $n_e (R_B)=0.02 \cm^{-3}$,
and the temperature is $T(R_B)=3.5 \times 10^6 \K$ \citep{Wong2011}.
The Bondi radius is given by
\begin{equation}
R_B \simeq \frac{2 G M_{\rm BH}}{c_s^2} = 220
\left( \frac {M_{\rm BH}} {2 \times 10^9 M_\odot} \right)
\left( \frac {T} {3.5 \times 10^6 \K } \right) \pc,
\label{eq:rb1}
\end{equation}
where $c_s$ is the sound speed in the undisturbed gas.
The temperature and electron density increase inward, reaching values of
$T_{20} \simeq 10^7 \K$ and $n_{e20} \simeq 0.3 \cm^{-3}$ at $r=20 \pc$ (\citealt{Wong2011}; no values are given at smaller radii).
We also note that in NGC~3115 the BH gravity dominates that of the galaxy to a distance of $\sim 30 \pc$ as the
black hole mass is $M_{\rm BH}=2 \times 10^9 M_\odot$.

\begin{figure}[htb]
\begin{center}
\includegraphics[width=0.75\textwidth]{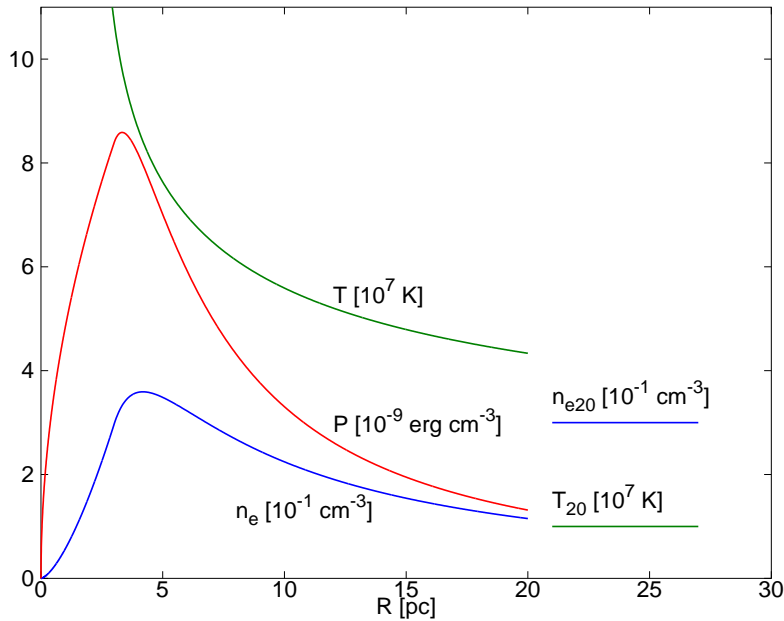}
\caption{The average density and pressure of the hot bubble according to equations (\ref{eq:rho1}) and (\ref{eq:pe2}),
as well as the temperature that is calculated from the pressure for the stellar density profile given in equation \ref{eq:nast1}.
The escape velocity parameter is $\beta=1$ (eq. \ref{eq:pe11}), and the mass loss parameter of $\eta=0.1$ (eq. \ref{eq:ramp1}) is taken
to crudely fit the ISM properties of NGC~3115 at $r=20 \pc$, {{{ shown in the figure as the horizontal lines}}}.
 }
\label{fig:Pr2}
\end{center}
\end{figure}

The average density and pressure of the hot bubble according to equations (\ref{eq:rho1}) and (\ref{eq:pe2}),
are drawn in Fig. \ref{fig:Pr2} for a SMBH mass of $M_{\rm BH}=2 \times 10^9 M_\odot$, and a stellar density given by
\begin{equation}
n_\ast =  5\times 10^5 \pc^{-3}
\begin{cases}
1, & r \le 3 \pc \\
(r / 3 \pc)^{-3}, & r > 3 \pc,
\end{cases}
\label{eq:nast1}
\end{equation}
and for $\beta=1$ (eq. \ref{eq:pe11}) and $\eta=0.1$ (eq. \ref{eq:ramp1}).
The density within $r = 3 \pc$ is from \cite{Kormendy1996}, while at $r > 3 \pc$ is our assumption.
{{{ The particular form of the decline in stellar density at $r>3 \pc$ has no significant consequences,
and the particular power law was chosen for the sake of simplicity and definite calculations. }}}
The value of the mass loss efficiency, which is the fraction of the mass lost by stars that ends up as hot gas
in the hot bubble, is chosen as $\eta=0.1$ to more or less match the pressure and density of the ISM at $r=20 \pc$.
It is a parameter of the model that should be typically in the range of $\sim 0.1-1$.
The temperature that is calculated from the pressure is also drawn on Fig. \ref{fig:Pr2}.
Beyond $\sim 30 \pc$ the average temperature is only $\sim 2$ times as large as the virial temperature of the cluster,
and our assumptions of a hot bubble become inadequate.
% FFFFFFFFFFFFFFFFFFFFFFFFFFFFFFFFFFFFFFFFFFFFFFFF
% FFFFFFFFFFFFFFFFFFFFFFFFFFFFFFFFFFFFFFFFFFFFFFFF

The following conclusions emerge from Fig. \ref{fig:Pr2}.
(1) The pressure of the shocked stellar winds of the high-velocity circum-SMBH stars is larger than the
ISM pressure near the center, even for a mass loss efficiency of only $\eta \sim 0.1$.
This accounts, we argue, for the accretion rate of NGC~3115 being much lower than the Bondi accretion rate \citep{Wong2011}.
(2) At the center, $r < 3 \pc$, the rate of mass loss into the hot gas per unit volume is
$\dot \chi \equiv (n_\ast \eta \dot m_\ast)_c = 5 \times 10^{-6} M_\odot \pc^{-3} \yr^{-1}$.
Even if this value is ten times lower, a hot bubble with pressure larger than the ISM pressure of NGC~3115 can still be formed.
(3) For $\dot \chi \la 10^{-5} M_\odot \pc^{-3} \yr^{-1}$ the hot bubble's density is lower than that of the ISM.
This structure is Rayleigh-Taylor (RT) unstable. This structure is analyzed below.

{{{ We note that the structure presented here is a temporary one. Eventually, the gas in the center originated from stellar winds
will radiatively cool and form cold clumps. Some will be accreted and amplify the AGN activity.
Many other clumps will be evaporated by the hot bubble and by the new AGN activity.
Accretion of clumps onto a SMBH in a turbulent medium was studied by \cite{Hobbs2011}, and accretion of cold clumps onto
Sgr A$^{\ast}$ in the Milky Way was simulated by \cite{Cuadra2008}. }}}

%%%
%%%    The gas pressure is
%%%    \begin{equation}
%%%    P_g=2.6 \times 10^{-10}
%%%    \left( \frac {n_e} {0.1 \cm^{-3}} \right)
%%%    \left( \frac {T} {10^7 \K } \right) \erg \cm^{-3}.
%%%    \label{eq:pgas1}
%%%    \end{equation}

% ==========================================================
\section{A TENUOUS HOT BUBBLE FORMED BY STELLAR OR AGN WINDS}
\label{s-hotbubble}
% ==========================================================

 We found above that in some cases the hot bubble that formed by the stellar winds of circum-SMBH high-velocity stars
can have a lower density than the ISM while its pressure is about equal to the ISM pressure $P_{\rm ISM}$.
This situation is prone to RT instability.
The same might hold for AGN winds.
The power of the winds that is required to form a hot bubble that can support the ISM is
\begin{equation}
W_{\rm wind} \simeq \frac{3}{2}P_{\rm ISM} V \tau_{\rm esc}^{-1}
= 3 \times 10^{37}  \beta
\left( \frac{T n_e}{10^7 \K \cm^{-3}} \right)
\left( \frac {M_{\rm BH}}{10^9 M_\odot} \right)^{1/2}
\left( \frac {R_h} {1 \pc} \right)^{3/2} \erg \s^{-1}
\label{eq:w1},
\end{equation}
where the escape time $\tau_{\rm esc}$ is given by equation (\ref{eq:pe11}), and $V$ is the volume of the hot bubble.
This implies that even a very weak AGN wind can form such a bubble.
With an efficiency of $1 \%$, namely, $W_{\rm wind}=0.01 \dot M_{\rm BH} c^2$, the required accretion rate
is $\dot M_{\rm BH} = 5 \times 10^{-8} M_\odot \yr^{-1}$.
For comparison, we note that the Chandra upper limit on the luminosity of NGC~3115 is $ \sim 10^{38} \erg \s^{-1}$ \citep{Diehl2008},
{{{ and the radio power is $L_{\rm radio}=3 \times 10^{35} \erg \s^{-1}$ \citep{Wrobel2012}.
The luminosity of the hot bubble as studied here has a low X-ray luminosity compared with the external gas.
First, the volume of the bubble is very small.
Second, the density inside the bubble is lower than that of the surrounding gas, and hence its
emissivity is lower. Even if more of the stellar wind incorporated to the bubble, the X-ray luminosity
from the bubble is much below detection limits.
}}}

The flow structure considered in this section has the following properties.
The hot bubble is continuously supplied by hot gas from the shocked stellar winds or the AGN wind or jets.
A pressure equilibrium is maintained between the hot bubble and the ISM, and a
structure of a hot tenuous gas supporting a denser and cooler gas is achieved.
This structure is RT unstable.
Such a structure, we claim, is similar to the density inversion found in the outer atmosphere of
red giant stars (e.g., \citealt{Harpaz1984, Freytag2008}), but not identical.
At the outer edge of the recombination zone of hydrogen in red giant stars the convection heat transfer
becomes less efficient.
The requirement to transfer energy leads to a steep temperature gradient that in turn causes a density inversion,
i.e., the density increases outward (e.g., \citealt{Harpaz1984}).
This occurs in the convective region, which is already unstable.
In the density-inversion layer in stars, therefore, cold convective cells fall and hot convective cells
buoy outward.
We suggest that the same process occurs in the flow structure discussed here.

There are some basic differences in the properties of the density-inversion layers of stars and of the case studied here.
The main differences are that the hot gas in our case buoys to large distances, and fresh gas from
stellar wind or the AGN replaces it.
Also, the entire region is optically thin, unlike stars where it is optically thick.
In stars the width of the density-inversion region is determined by heat transfer requirements, whereas in our
case it is determined by dynamics, mixing, and local heat conduction.
In stars the density scale height is not much shorter than the pressure scale height $l_p$.
The size of the convective cells is taken to be of the order of the pressure scale height.
In our case the density can change by an order of magnitude from the inner tenuous region to the denser outer ISM,
and we expect the RT instability to break the cells to smaller cells.
We therefore take the size of the rising and falling gas elements to be $R_c \ll l_p$.

We take the density-inversion zone to be of the order of the pressure scale height (in stars it can be much smaller).
For a central gravity source the pressure scale height for a constant temperature is given by
\begin{equation}
l_p = {R_h} \left[ \frac{C_i}{u_{\ast} (R_h)} \right]^2,
\label{ee:lp1}
\end{equation}
where $C_i$ is {{{ the}}} isothermal sound speed,
{{{ and $u_\ast(R_h)$ is the stellar velocity given in equation (\ref{eq:vbh1}) and evaluated at the radius of the hot bubble $R_h$.
The shocked stellar wind will be heated to a temperature of $T \approx (3/16) m u_{\ast}^2/k$,
where $m$ is the mean mass per particle in the gas. The sound speed is
$[(5/3) kT/m]^{1/2} \approx 0.6 u_{\ast}$. Thus, we can take $l_p \sim R_h$.}}}
Therefore, we assume first that the width of the density-inversion layer is $\Delta r_i \sim R_h$.

Consider then a spherical parcel of gas (a blob) of radius $R_c$ and density of $\rho_c$ moving with a terminal
velocity $v_b$ through an external medium of density $\rho_e$.
The buoyancy force on the blob is
\begin{equation}
 F_b = \left( \rho_e - \rho_c \right) \frac{4}{3} \pi R_c^3 g,
 \label{ee:vt1}
\end{equation}
{{{ where $g$ is the gravitational acceleration.}}} The drag force on the bubble is
\begin{equation}
 F_d \approx \frac{1}{2} C_D \pi R_c^2 \rho_e v_t^2,
  \label{ee:vt2}
\end{equation}
where $C_D \simeq 0.75$ (\citealt{Kaiser2003}).
Assuming $\rho_c \ll \rho_e$ and taking $g = u_\ast^2 / R_h$, the terminal velocity of the bubble is
\begin{equation}
 v_t \approx \left( {\frac{8}{3 C_D}} \right)^{1/2}
 \left( \frac{R_c}{R_h} \right)^{1/2} u_\ast = \beta u_\ast,
  \label{ee:vt3}
\end{equation}
where in the second equality we identify the terminal velocity as the velocity by which the hot gas {{{ escapes}}} from
the hot bubble outward, with
\begin{equation}
\beta \simeq  0.6 \left( \frac{R_c}{0.1R_h} \right)^{1/2} .
  \label{ee:beta1}
\end{equation}

Complex processes take place in the density-inversion layer.
(1) Heat conduction time scale over a distance of $\Delta r_T = R_c \sim 0.1 \pc$ and a temperature difference of $\Delta T=10^7 \K$,
is few$\times 10 \yr$. This is shorter than the fall time of a dense clump from $\sim 1 \pc$. Therefore, the hot bubble gas heats the
clump by heat conduction. Closer to the center, the clump will be shredded to smaller cells. Hence, before the dense ISM clumps can reach the center
{{{ they}}} will be evaporated. This is not true for denser and cooler blobs that fall inward, as in the cold feedback mechanism \citep{Pizzolato2005}.
(2) Because of the stellar motion and/or AGN activity, the density-inversion layer is expected to be more chaotic than just a RT-unstable region.
There will be vortices that will increase mixing, namely, {{{ reduce}}} the effective value of $\Delta r_T$.

%
% ==========================================================
\section{DISCUSSION AND SUMMARY}
\label{s-summary}
% ==========================================================

We studied the pressure exerted by the winds of circum-SMBH high-velocity stars on the surrounding ISM.
We found that in some cases this pressure is significant and can substantially suppress the inflow of the
ISM relative to what a simple Bondi accretion would give.
Our result can explain the finding of \citet{Wong2011} that the Bondi accretion rate calculated by them from
the ISM density and temperature is six orders of magnitude above the observed upper limit on the accretion rate in the S0 galaxy NGC~3115.

In section \ref{s-ngc3115} we quantitatively examined the situation in the galaxy NGC~3115.
Shocked winds of circum-SMBH high-velocity stars form a bubble of hot gas whose pressure is significant, as evident from
Fig. \ref{fig:Pr2}.
{{{ The colliding winds heat up to very high temperatures, build significant pressure, and are not expected to be accreted by the SMBH
even though they lose angular momentum.
Cooler clumps that fall inward, from the ISM or from inhomogeneities within the hot bubble, will encounter the winds
of fast-moving stars very close to the SMBH. This collision will heat such clumps, suppressing their accretion.
Even if there is a small accretion rate, a very weak disc wind from the accretion disc might
further lower the accretion rate. The study of the interaction of AGN winds with the gas
near the SMBH is a subject of a future study using numerical simulations.
}}}

There are some uncertainties in the model, such as the exact behavior of the stellar mass loss,
trajectories of stars around the SMBH, and the stochastic behavior of the post-shock stellar winds.
Some of these will be studied in future numerical simulations. However, the result that the stellar winds cannot be
ignored is robust.

For some values of the parameters we found that a situation might arise where the hot bubble's density is lower than the ISM density.
In this case, Rayleigh-Taylor (RT) instability takes place, and a density-inversion layer is formed (see schematic description in Fig. \ref{fig:fig1}).
Although hot tenuous gas buoys outward and dense ISM gas moves inward, the density-inversion layer itself continues to exist.
The ISM gas is heated near the center and accumulated {{{ into}}} the hot bubble.

{{{ While the scenario suggested here may explain the low X-ray luminosity observed in the galaxy NGC~3115,
its properties have not yet been observed or affirmed directly.
The size of the hot bubble described is below the resolution limit of the observations and cannot yet be observed.
Alternative explanations for a below-Bondi accretion rate are mentioned in section \ref{s-intro}.
}}}

{{{ We note that in our scenario there can be no steady state over a very long time of $ \sim 10^7-10^8 \yr$.
Over this time scale radiative cooling becomes important and more of the cooling gas will be accreted by the SMBH.
This will lead to stronger AGN activity that will heat and expel gas, hence reducing back the accretion rate and AGN power.
In addition stellar formation must occur from time to time.
Most likely, there are local star-burst episodes when the accretion rate is much higher than the Bondi accretion rate.
The high accretion rate is probably driven by cold clumps (filaments, streams).
Indeed, the stellar-wind pressure cannot prevent accretion of very dense clouds.}}}

Our result is more general in showing that in many cases the Bondi accretion process does not work because
one of its basic assumptions, that there is no central pressure, breaks down.
This is one of several reasons why the Bondi accretion model may not apply in some cases (see section \ref{s-intro}).

Finally, we note that our model may be relevant for active galaxies where the hot bubble might be formed by the AGN jets or winds.
For typical values of AGN jets and winds the hot bubble density will be low, and a density-inversion layer will be formed.
We expect this process to be of high significance in the process of AGN feedback acting in young galaxies.
Barring Bondi-like accretion, dense and cold clumps in the ISM can still flow inward and feed the SMBH.
Namely, AGN feedback mechanisms require the feeding to be by cold clumps, i.e., a cold feedback mechanism.

{{{ We thank an anonymous referee for many detail and very helpful comments that substantially improved the manuscript. }}}
This research was supported by the Asher Fund for Space Research and the E. and J. Bishop Research Fund at the Technion,
and the Israel Science foundation.

%   %%%%%%%%%%%%%%%%%%%%%%%%%%%%%%%%%%%%%%%%%%%%%%%%% CLOSURE
%\clearpage

\end{document}